\documentclass[times,final,twocolumn]{IEEEtran}

\usepackage{amsmath}
\usepackage{xcolor}
\usepackage[final]{graphicx}
\usepackage{epstopdf}
\usepackage{subcaption}
\usepackage{multirow}
\usepackage{url}
\usepackage{makecell}
\usepackage{hyperref}
\usepackage[normalem]{ulem}
\usepackage[inline]{enumitem}

\usepackage[noadjust]{cite} 

\definecolor{usc}{rgb}{0.6,0.106,0.117}
\usepackage{color}

\newcommand{\eg}{\textit{e.g.,\ }}

\newcommand{\ie}{\textit{i.e.,\ }}
\newcommand{\etal}{\textit{et al.\@}}

\begin{document}
	
\title{Neural Predictive Coding using Convolutional Neural Networks towards Unsupervised Learning of Speaker Characteristics}

\author{Arindam~Jati,
	and~Panayiotis~Georgiou,~\IEEEmembership{Senior~Member,~IEEE}}

\maketitle
\begin{abstract}
  Learning speaker-specific features is vital in many applications like speaker recognition, diarization and speech recognition.
  This paper provides a novel approach, we term Neural Predictive Coding (NPC), to learn speaker-specific characteristics in a completely unsupervised manner from large amounts of unlabeled training data that even contain many non-speech events and multi-speaker audio streams.
  The NPC framework exploits the proposed short-term active-speaker stationarity hypothesis which assumes two temporally-close short speech segments  belong to the same speaker, and thus a common representation that can encode the commonalities of both the segments, should capture the vocal characteristics of that speaker. 
  We train a convolutional deep siamese network to produce ``speaker embeddings'' by learning to separate `same' vs `different' speaker pairs which are generated from an unlabeled data of audio streams.
  Two sets of experiments are done in different scenarios to evaluate the strength of NPC embeddings and compare with state-of-the-art in-domain supervised methods.
  First, two speaker identification experiments with different context lengths are performed in a scenario with comparatively limited within-speaker channel variability.
  NPC embeddings are found to perform the best at short duration experiment, and they provide complementary information to i-vectors for full utterance experiments. 
  Second, a large scale speaker verification task having a wide range of within-speaker channel variability is adopted as an upper-bound experiment where comparisons are drawn with in-domain supervised methods.
\end{abstract}

\begin{IEEEkeywords}
Speaker-specific characteristics, unsupervised learning, Convolutional Neural Networks (CNN), siamese network, speaker recognition.
\end{IEEEkeywords}

\IEEEpeerreviewmaketitle

\section{Introduction}
\label{sec:intro}

\IEEEPARstart{A}{coustic} modeling of speaker characteristics is an important task for many speech-related applications. It is also a very challenging problem due to the highly complex information that the speech signal modulates, from lexical content to  emotional and behavioral attributes~\cite{Huang:2001:SLP:560905, narayanan2013behavioral} and multi-rate encoding of this information.
A major step towards speaker modeling is to identify features that focus only on the  speaker-specific characteristics of the speech signal. Learning these characteristics has various applications in speaker segmentation~\cite{kotti2008speaker}, diarization~\cite{anguera2012speaker}, verification~\cite{chen2003towards}, and recognition~\cite{campbell1997speaker}. 
State-of-the-art methods for most of these applications use short-term acoustic features~\cite{hansen2015speaker} like MFCC~\cite{davis1980comparison} or PLP~\cite{hermansky1990perceptual} for signal parameterization. In spite of the effectiveness of the algorithms used for building speaker models~\cite{campbell1997speaker} or clustering speech segments~\cite{anguera2012speaker}, sometimes these features fail to produce high between-speaker variability and low within-speaker variability~\cite{hansen2015speaker}. 
This is because MFCCs contain a lot of supplementary information like phoneme characteristics, and they are frequently deployed in speech recognition~\cite{rabiner1993fundamentals}.

\subsection{Prior work}
Significant research effort has gone into solving the above mentioned discrepancies of short-term features by incorporating long-term or prosodic features~\cite{shriberg2005modeling} into existing systems. 
These features can specifically be used in speaker recognition or verification systems since they are calculated at utterance-level~\cite{hansen2015speaker}. 
Another way to tackle the problem is to calculate  mathematical functionals or transformations on top of MFCC features to expand the context and project them on a ``speaker space'' which is supposed to capture speaker-specific characteristics.
One popular method~\cite{reynolds2000speaker} is to build a GMM-UBM~\cite{hansen2015speaker} on training data and utilize MAP adapted GMM supervectors~\cite{reynolds2000speaker} as fixed dimensional representations of variable length utterances.     
Along this  line of research, there has been ample effort in exploring different factor analysis techniques on the high dimensional supervectors to estimate contributions of different latent factors like speaker- and channel-dependent variabilities \cite{kenny2004disentangling}.
Eigenvoice and eigenchannel methods were proposed by Kenny \etal~\cite{kenny2003new} to separately determine the contributions of speaker and channel variabilities respectively. 
In 2007, Joint Factor Analysis (JFA) \cite{kenny2007joint} was proposed to model speaker variabilities and compensate for channel variabilities, and it outperformed the former technique in capturing speaker characteristics.

{\color{usc}\textbf{\textit{Introduction of i-vectors:}}} 
In 2011, Dehak \etal~proposed i-vectors~\cite{dehak2011front} for speaker verification.
The i-vectors were inspired by JFA, but unlike JFA, the i-vector approach trains a unified model for speaker and channel variability. 
One inspiration for proposing the Total Variability Space~\cite{dehak2011front} was from the observation that the channel effects obtained by JFA also had speaker factors.
The i-vectors have been used by many researchers for numerous applications including speaker recognition~\cite{dehak2011front, dehak2009support}, diarization~\cite{sell2014speaker, dupuy2012vectors} and speaker adaptation during speech recognition~\cite{saon2013speaker} due to their state-of-the-art performance.    
But, performance of i-vector systems tends to deteriorate as the utterance length decreases \cite{kanagasundaram2011vector}, especially when there is a mismatch between the lengths of training and test utterances. 	
Also, the i-vector modeling, similar to most factor analysis methods, is constrained by the GMM assumption which might degrade the performance in some cases \cite{hansen2015speaker}. 

{\color{usc}\textbf{\textit{DNN-based methods in speaker characteristics learning:}}} 
Recently, Deep Neural Network- (DNN)~\cite{Goodfellow-et-al-2016} derived ``speaker embeddings''~\cite{rouvier2015speaker} or bottleneck features~\cite{hinton2006reducing} have been found to be very powerful for capturing speaker characteristics.
For example, in~\cite{yamada2013improvement, variani2014deep} and ~\cite{ghalehjegh2015deep}, frame-level bottleneck features have been extracted using DNNs trained in a supervised fashion over a finite set of speakers; and some aggregation techniques like GMM-UBM~\cite{reynolds2000speaker} or i-vectors have been used on top of the frame-level features for utterance-level speaker verification. Chen \etal~\cite{chen2011learning, chen2011extracting} developed a deep neural architecture and trained it for frame-level speaker comparison task in a supervised way. They achieved promising results in speaker verification and segmentation tasks even when they evaluated their system on out-of-domain data~\cite{chen2011learning}. In~\cite{snyder2016deep}, the authors proposed an end-to-end text-independent speaker verification method using DNN embeddings. It uses the similar approach to generate the embeddings, but the utterance-level statistics are computed internally by a pooling layer in the DNN architecture.      
In more recent work~\cite{li2017deep}, different combinations of Convolutional Neural Networks (CNN) and Recurrent Neural Networks (RNN)~\cite{Goodfellow-et-al-2016} have been exploited to find speaker embeddings using the triplet loss function which minimizes intra-speaker distance and maximizes inter-speaker distance~\cite{li2017deep}. 
The model also incorporates a pooling and normalization layer to produce utterance-level speaker embeddings.

{\color{usc}\textbf{\textit{Need for unsupervised methods and existing works:}}} 
In spite of the wide range of DNN variants, all these need one or more annotated dataset(s) for supervised training. 
This limits the learning power of the methods, especially given the data-hungry needs of advanced neural network-based models. 
Supervised training can also  limit robustness due to  over-tuning to the specific training environment. 
This can cause degradation in performance if the testing condition is very different from that of the training. 
Moreover, transfer learning~\cite{pan2010survey} of the supervised models to a new domain also needs labeled data.  
This points to a desire and opportunity in employing unlabeled data and unsupervised methods for learning speaker embeddings.

There have been a few efforts~\cite{saleem2016discriminative, zhang2015multilayer, lapidot2002unsupervised} in the past to employ neural networks for acoustic space division, but these works focused on speaker clustering and they did not exploit short-term stationarity towards embedding learning.
In~\cite{lee2009unsupervised}, an unsupervised training scheme using convolutional deep belief networks has been proposed for audio feature learning. They applied those features for phoneme, speaker, gender and music classification tasks. Although, the training employed there was unsupervised, the proposed system for speaker classification was trained on TIMIT dataset~\cite{timit_data} where every utterance is guaranteed to come from a single speaker, and PCA whitening was applied on the spectrogram per utterance basis. Moreover, performance of the system on out-of-domain data was not evaluated. 

\subsection{Proposed work}
In this paper, we propose a completely unsupervised method for learning features having speaker-specific characteristics from unlabeled audio streams that contain many non-speech events (music, noise, and anything else available on YouTube). We term the general learning of signal characteristics  via the short-term stationarity assumption  \emph{Neural Predictive Coding} (NPC) since it was inspired by the idea of predicting present value of signal from a past context as done in Linear Predictive Coding (LPC)~\cite{o1988linear}.
The short-term stationarity assumption can take place, according to the frame size, along different characteristics. For example we can assume that the behaviors expressed in the signal will be mostly stationary within a window of a few seconds as we did in \cite{li2016_unsupervised-la}. In this work we assume that any potentially active speaker will be mostly stationary within a short window: the active speaker is unlikely to change multiple times within a couple of seconds.
LPC predicts future values from past data via a filter described by its coefficients. NPC can predict future values from past data via neural network. The embedding inside the NPC neural network can serve as a feature.
Moreover, while predicting future values from past, the NPC model can incorporate knowledge learned from big, unlabeled datasets.

The short-term speaker stationarity assumption was exploited in our previous work~\cite{jati2017_Speaker2Vec} via an encoder-decoder model to predict future values from past through a bottleneck layer.
The training involved in that work was able to see past and future values of the signal only from the `same speaker', assuming speaker stationarity. 
In contrast, the currently employed siamese architecture~\cite{chopra2005learning, hadsell2006dimensionality} helps the model to encounter and compare whether a pair of speech segments come from the same speaker or, two different speakers, based on unlabeled data via the short-term stationarity assumption.

We perform experiments under different scenarios and for different applications to explore the ability of the proposed method to learn speaker characteristics.
Moreover, the NPC training is done on out-of-domain data, and its performance is compared with i-vectors and recently introduced x-vectors~\cite{snyder2018x} trained on in-domain data.

Note that the NPC training needs no labels at all, not even speaker homogeneous regions.
For that reason, we do not expect NPC-derived features to beat in-domain supervised algorithms, but rather present this as an upper-bound aim.

The comparison reveals interesting directions that can arise through further introduction of context.
For example, if the algorithm employs longer same-speaker context (than 2\textit{s} assumed by this work) similar to i-vector systems then it can allow for variable length features and increased channel normalization.

Below are  the major aims of the proposed work towards establishing a robust speaker embedding:
\begin{enumerate*}
\item Training should require no labels of any kind (no speaker id labels, or speaker homogeneous utterances for training);
\item System should be highly scalable relying on  plentiful availability of unlabeled data;
\item Embedding should represent short-term characteristics and be suitable as an alternative to MFCCs in an aggregation system like \cite{yamada2013improvement} or ~\cite{ghalehjegh2015deep}; and, 
\item The training scheme should be readily applicable for unsupervised transfer learning.
\end{enumerate*}

The rest of the paper is organized as follows. 
The NPC methodology is described in Section~\ref{sec:NPC}. 
Section~\ref{sec:exp_sett} provides details about evaluation methodology and required experimental setup. 
Results are tabulated and discussed in Section~\ref{sec:results}. 
A qualitative analysis and future scopes are provided in Section~\ref{sec:Discussion_FutureWork}.
Finally conclusions are drawn in Section~\ref{sec:conclusion}.

\begin{figure*}[t]
	\centering
	\includegraphics[trim=5 220 30 45, clip,width=\textwidth]{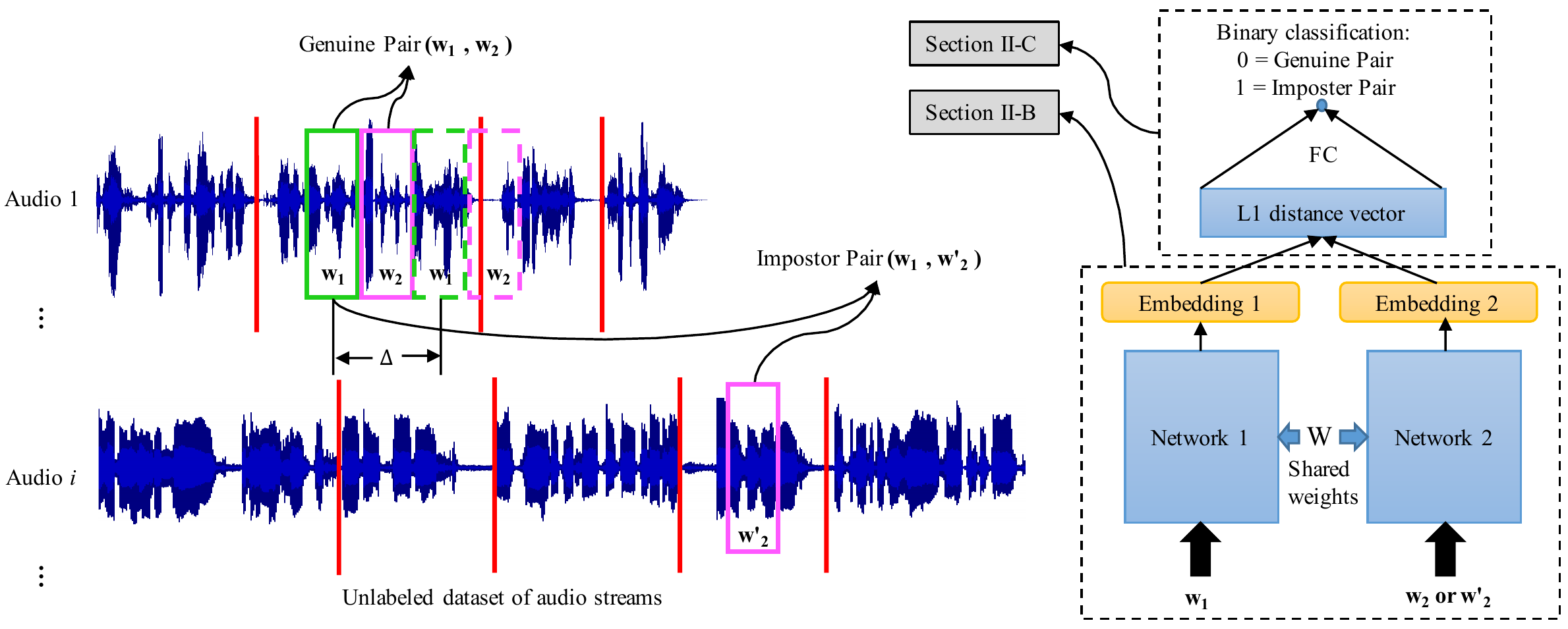}
	\caption{ NPC training scheme utilizing short-term speaker stationarity hypothesis. \textbf{Left:} Contrastive sample creation from unlabeled dataset of audio streams. Genuine and impostor pairs are created from unlabeled dataset as explained in Section~\ref{sec:Contrastive sample creation}. \textbf{Right:} The siamese network training method. The genuine and impostors pairs are fed into it for binary classification. ``FC'' denotes Fully Connected hidden layer in the DNN. Note that the siamese convolutional layers have been discussed in Section~\ref{sec:cnn}, and the derivation of the loss functions by comparing the siamese embeddings has been shown in Section~\ref{sec:siamese_dnn}.}
	\label{fig:NPC}
\end{figure*}

\section{Neural Predictive Coding (NPC) of Speaker Characteristics}
\label{sec:NPC}
Our ultimate goal is to learn a non-linear mapping (the employed DNN or part of it) that can project a small window of speech from any speaker to a lower dimensional embedding space where it will retain the maximum possible speaker-specific characteristics and reject other information as much as possible.

We expect that based on the unsupervised training paradigm we employ the embedding may also capture additional information, mainly channel characteristics and we intend to address that in future work, as further discussed in Section~\ref{sec:Discussion_FutureWork}.

\subsection{Contrastive sample creation}
\label{sec:Contrastive sample creation}
NPC learns to extract speaker characteristics in a contrastive way \textit{i.e.} by distinguishing between different speakers. During training phase, it possesses no information about the actual speaker identities, but only learns whether two input audio chunks are generated from the same speaker or not. 
We provide the NPC model two kinds of samples~\cite{chopra2005learning}. The first kind consists of pairs of speech segments that come from the same speaker, called \textit{genuine pairs}. The second type consists of speech segments from two different speakers, called \textit{impostor pairs}. This approach has been used in the past for numerous applications~\cite{hadsell2006dimensionality, chopra2005learning, koch2015siamese}, but all of them needed labeled datasets. The challenge is how we can create such samples if we do \textit{not} have labeled acoustic data. We exploit the characteristics of speaker-turntaking that result in \textit{short-term speaker stationarity} \cite{jati2017_Speaker2Vec}. The hypothesis of short-term speaker stationarity is based on the notion that given a long observation of human interaction, the probability of fast speaker changes will be at the tails of the distribution. In short: it is very unlikely to have extremely fast speaker changes (for example every 1~second). So, if we take pairs of consecutive \emph{short} segments from such a long audio stream, most of the pairs will contain two audio segments from the same speaker (genuine pairs). There will be definitely some pairs containing segments from two different speakers, but number of such pairs will probably be small compared to the total number of genuine pairs. To find the impostor pairs, we choose two random segments from two different audio streams in our unsupervised dataset. Again, intuitively the probability of finding the same speaker in an impostor pair is relatively lower than the probability of getting two different speakers in it, provided a sufficiently large unsupervised dataset. For example, sampling two random YouTube videos, the likelihood of getting the same speaker in both is very low.

The left part of Fig.~\ref{fig:NPC} shows this contrastive sample creation process. Audio stream $1$ and audio stream $i$ (for any $i$ between 2 to $N$, where $N$ is the number of audio streams in the dataset) are shown here. Assume the vertical red lines denote (unknown) speaker change points. $(\mathbf{w_1}, \mathbf{w_2})$ is a window pair where each of the two windows has $d$ feature frames. This window pair is moved over the audio streams with a shift of $\Delta$ to get the genuine pairs. For every $\mathbf{w_1}$, we randomly pick a window $\mathbf{w_2'}$ of same length from a different audio stream to get an impostor pair. All these samples are then fed into the siamese DNN network for binary classification of whether an input pair is genuine or impostor.

A siamese neural network (please see right part of Fig.~\ref{fig:NPC}), first introduced for signature verification~\cite{bromley1994signature}, consists of two identical twin networks with shared weights that are joined at the top by some energy function~\cite{hadsell2006dimensionality, chopra2005learning, koch2015siamese}. Generally, the siamese networks are provided with two inputs and trained by minimizing the energy function which is a predefined metric between the highest level feature transformations of both the inputs. The weight sharing ensures similar inputs are transformed into embeddings in close proximity with each other. The inherent structure of a siamese network enables us to learn similar or dissimilar input pairs with discriminative energy functions~\cite{chopra2005learning, hadsell2006dimensionality}. Similar to~\cite{koch2015siamese}, we use $L_1$ distance loss between the highest level outputs of the siamese twin networks for the two inputs. 

We will first describe in Section~\ref{sec:cnn} about the CNN that processes the speech spectrogram to automatically learn features to generate the embeddings.
Next, in Section~\ref{sec:siamese_dnn} we will discuss about the top part of the neural network of Fig.~\ref{fig:NPC} that involves comparing the two embeddings and deriving the final output and error for back-propagation.

\subsection{Siamese  Convolutional layers}
\label{sec:cnn}
The amazing effectiveness of CNNs have been well established in computer vision field~\cite{krizhevsky2012imagenet, simonyan2014very}. Recently, speech scientists are also applying CNNs for different challenging tasks like speech recognition~\cite{abdel2014convolutional,hinton2012deep}, speaker recognition~\cite{li2017deep, mclaren2014application,lukic2016speaker}, large scale audio classification~\cite{DBLP:journals/corr/HersheyCEGJMPPS16} \textit{etc.} The general benefits of using CNNs can be found in~\cite{Goodfellow-et-al-2016} and in the above papers. In our work, the inspiration to use CNNs comes from the need of exploring spectral and temporal contexts together through 2D convolution over the mel-spectrogram features (please see Section~\ref{sec:Feature extraction} for more information). 
The benefits of such a 2D convolution have also been shown with more traditional signal processing feature sets such as Gabor features \cite{chang2014robust}.

\begin{figure*}[t]
	\centering
	\includegraphics[trim=20 165 180 150, clip,width=\textwidth]{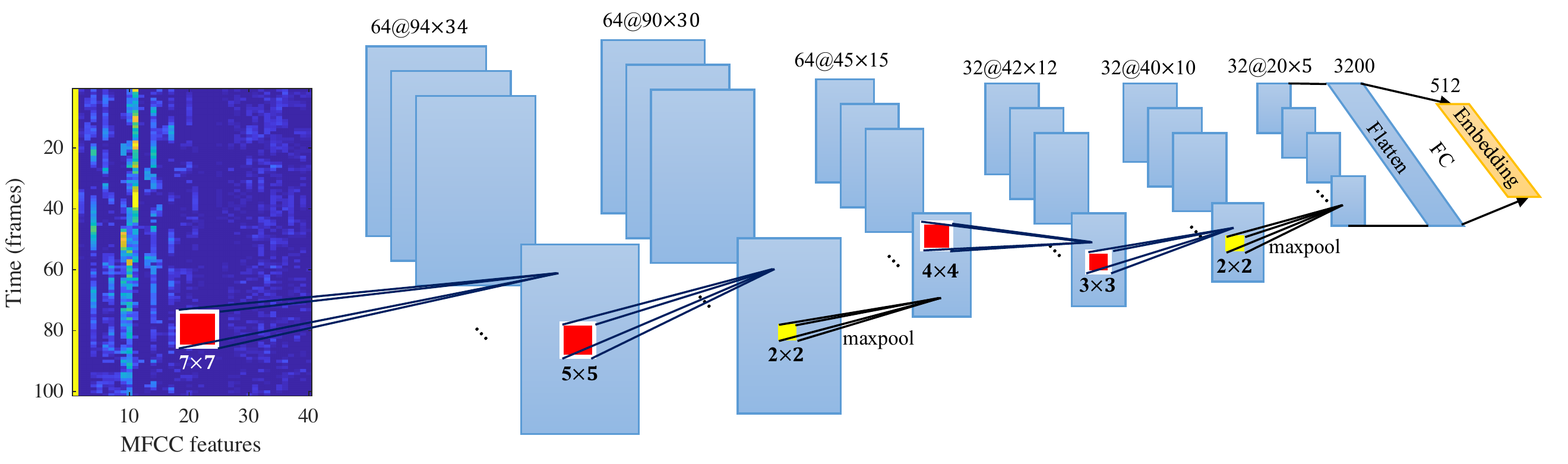}
	\caption{The DNN architecture employed in each of the siamese twins. All the weights are shared between the twins. The kernel sizes are denoted under the red squares. $2\times2$ max-pooling is used as shown by yellow squares. All the feature maps are denoted as: $N@x\times y$, where $N=\text{number of feature maps}$, $x\times y=\text{size of each feature map}$. Dimension of the speaker embedding is 512. ``FC'' = Fully Connected layer.}
	\label{fig:DNN}
\end{figure*}

Our siamese network (one of the identical twins), built using multiple CNN layers and one dense layer at the highest level, is shown in Fig.~\ref{fig:DNN}. 
We gradually reduce the kernel size from $7\times7$ to $5\times5$, $4\times4$, and $3\times3$. We have used $2\times2$ max-pooling layers after every two convolutional layers. The size of stride for all convolution and max-pooling operations has been chosen to be 1. We have used Leaky ReLU nonlinearity~\cite{maas2013rectifier} after every convolutional or fully connected layer (omitted from  Fig.~\ref{fig:DNN} for clearer visualization). 

We have applied batch normalization~\cite{pmlr-v37-ioffe15} after every layer to reduce the ``internal covariance shift''~\cite{pmlr-v37-ioffe15} of the network. It helped the network to avoid overfitting and converge faster without the need of using dropout layers~\cite{srivastava2014dropout}.
After the last convolutional layer, we get 32 feature maps, each of size $20\times5$. We flatten these maps to get a $3200$ dimensional vector which is connected to the final $512$ dimensional NPC embedding through a fully connected layer. 
The embeddings are obtained before applying the Leaky ReLU non-linearity~\footnote{Following standard convention for extracting embedding from DNNs, such as in \cite{snyder2017deep}.}.

\subsection{Comparing Siamese embeddings -- Loss functions\label{sec:siamese_dnn}}

Let $\mathbf{f(x_1)}$ and $\mathbf{f(x_2)}$ be the highest level outputs of the siamese twin networks for inputs $\mathbf{x_1}$ and $\mathbf{x_2}$ (in other words, $\mathbf{(x_1,x_2)}$ is one contrastive sample obtained from the window pair $(\mathbf{w_1,w_2})$ or $(\mathbf{w_1,w_2'})$). We will use this transformation $\mathbf{f(x)}$ as our ``embedding'' for any input $\mathbf{x}$ (please see right part of Fig.~\ref{fig:NPC}). Here $\mathbf{x}$ is a matrix of size $d\times m$, and it denotes $d$ frames of $m$ dimensional MFCC feature vectors in window $\mathbf{w}$. Similarly, $\mathbf{x_i}$ denotes the feature frames in window $\mathbf{w_i}$ for $i=1,2$.
We have deployed two different types of loss functions for training the NPC model. They are described below.
\subsubsection{Cross entropy loss}
\label{sec:Cross entropy loss}
Inspired from~\cite{koch2015siamese}, the loss function is designed in a way such that it decreases the weighted $L_1$ distance between the embeddings $\mathbf{f(x_1)}$ and $\mathbf{f(x_2)}$ if $\mathbf{x_1}$ and $\mathbf{x_2}$ are from a genuine pair, and increases the same if they are from an impostor pair. 

The ``$L_1$ distance vector'' (Fig. \ref{fig:NPC}, right) is obtained by calculating element-wise absolute difference between the two embedding vectors $\mathbf{f(x_1)}$ and $\mathbf{f(x_2)}$ and is given by:
\begin{equation}
\textbf{L}(\mathbf{x_1},\mathbf{x_2})=\vert \mathbf{f(x_1)}-\mathbf{f(x_2)} \vert .
\end{equation}            
We connect $\mathbf{L(x_1,x_2)}$ to two outputs $g_i(\mathbf{x_1, x_2})$ using a fully connected layer:
\begin{equation}
\label{eqn:g(x1,x2)}
g_i(\mathbf{x_1}, \mathbf{x_2}) = \sum_{k=1}^{D}w_{i,k} \times \vert f(\mathbf{x_1})_k-f(\mathbf{x_2})_k \vert + b_i
\end{equation}
for $i=1,2$. Here, $f(\mathbf{x_1})_k$ and  $f(\mathbf{x_2})_k$ are the $k^{th}$ elements of $\mathbf{f(x_1)}$ and $\mathbf{f(x_2)}$ vectors respectively, and $D$ is the length of those vectors (so, D is the embedding dimension). 
$w_{i,k}$'s and $b_i$'s are the weights and bias for the $i^{th}$ output. 
Note that these weights and biases are affecting only the binary classifier, and they are not part of the siamese network.

A softmax layer produces the final probabilities:
\begin{align}
\label{eqn:p(x1,x2)}
p_i(\mathbf{x_1, x_2}) &= s(g_i(\mathbf{(x_1,x_2)})) && \text{for } i=1,2.
\end{align} 
Here $s(.)$ is the softmax function given by
\begin{align}
s(g_i\mathbf{(x_1,x_2)})&=\frac{e^{g_i(\mathbf{x_1,x_2})}}{e^{g_1(\mathbf{x_1,x_2})}+e^{g_2(\mathbf{x_1,x_2})}} &&\text{for } i=1,2.
\end{align}

The network in Fig.~\ref{fig:NPC} is provided with the genuine and impostor pairs as explained in Section \ref{sec:Contrastive sample creation}. 
We use cross entropy loss here. It is given by
\begin{align}
\begin{split}
e(\mathbf{x_1,x_2})&= 
-\mathcal{I}(y(\mathbf{x_1, x_2})=0) \log(p_1(\mathbf{x_1, x_2})) \\ 
&- \mathcal{I}(y(\mathbf{x_1, x_2})=1) \log(p_2(\mathbf{x_1, x_2}))
\end{split}
\end{align}
where $\mathcal{I}(.)$ is the indicator function defined as:
\begin{align*}
	\mathcal{I}(t)=
	\begin{cases}
	1, &\text{if }t\text{ is true}\\
	0, &\text{if }t\text{ is false}
	\end{cases}
\end{align*}
and, $y(\mathbf{x_1, x_2})$ is the true label for the sample $\mathbf{(x_1,x_2)}$, defined as:
\begin{equation}
\label{eqn:4}
y(\mathbf{x_1},\mathbf{x_2})=
\begin{cases}
0, &\text{if } (\mathbf{x_1} \text{, } \mathbf{x_2}) \text{ is a genuine pair.} \\
1, &\text{if } (\mathbf{x_1} \text{, } \mathbf{x_2}) \text{ is an impostor pair.} 
\end{cases}
\end{equation}
Using Equation~\ref{eqn:4}, we can write the error as
\begin{align}
\begin{split}
e(\mathbf{x_1,x_2})&= 
-(1-y(\mathbf{x_1, x_2})) \log(p_1(\mathbf{x_1, x_2})) \\ 
&- y(\mathbf{x_1, x_2}) \log(p_2(\mathbf{x_1, x_2}))
\end{split}
\end{align}

\subsubsection{Cosine embeddings loss}
\label{sec:Cosine embeddings loss}
We also analyze the performance of the network when we directly minimize a contrastive loss function between the embeddings. So, there is no need to add an extra fully connected layer at the end. The employed cosine embedding loss is defined below.

\begin{equation*}
\begin{aligned}
L_{cos}(\mathbf{x_1, x_2}) = 
\begin{cases}
1-C(\mathbf{f(x_1), f(x_2)}), & \text{if } y(\mathbf{x_1, x_2})=0 \\
C(\mathbf{f(x_1), f(x_2)}), &\text{if } y(\mathbf{x_1, x_2})=1
\end{cases} 
\end{aligned}
\end{equation*}
Here $C(\mathbf{f(x_1), f(x_2)})$ is the cosine similarity between $\mathbf{f(x_1)}$ and  $\mathbf{f(x_2)}$
defined as
\begin{equation*}
cos(\mathbf{x_1, x_2}) = \frac{\mathbf{x_1.x_2}}{||\mathbf{x_1}||_2||\mathbf{x_2}||_2}.
\end{equation*} 

Here $||.||_2$ denotes the $L_2$ norm. In Section~\ref{sec:results}, performances of the two types of loss functions will be analyzed through experimental evidence. 

\subsection{Extracting NPC embeddings for test audio}
Once the DNN model is trained, we use it for extracting speaker embeddings from any test audio stream. As discussed in Section~\ref{sec:siamese_dnn}, the transformation achieved by the siamese network on an input segment $\mathbf{x}$ of length $d$ frames is given by $\mathbf{f(x)}$. We use only this siamese part of the network to transform a sequence of MFCC frames of any speech segment into NPC embeddings by using a sliding window $\mathbf{w}$ of $d$ frames and shifting it by 1 frame along the entire sequence.   

\begin{table}[!t]
	\caption{NPC Training Datasets}
	\label{tab:table_datasets}
	\centering
	\begin{tabular}{c|c|c}
		\hline
		Name of the dataset & Size (hours) & Number of samples\\
		\hline
		Tedlium & 100 &358K \\
		Tedlium-Mix & 110 &395K\\
		YoUSCTube & 584 & 2.1M\\
		\hline
	\end{tabular}
\end{table}

\section{Evaluation Methodology}
\label{sec:exp_sett}
The nature of the proposed method introduces a great challenge in its evaluation. All existing speaker identification methods employ some level of supervision. For example x-vector systems \cite{snyder2018x} employ data with complete speaker labels, while i-vector systems \cite{dehak2011front} require labeling of speaker-homogeneous regions.

  In our proposed work we intend to establish a low-level speaker-specific feature, on which subsequent supervised methods or layers can operate.

  Given the above evaluation challenge we perform two sets of comparisons with existing methods:
\begin{enumerate}
\item \textbf{Speaker identification evaluation}: Speaker identification (\ie closed set multi-class speaker classification) experiments are performed at different context lengths. In that case we compare with other low-level features such as MFCCs and statistics of MFCCs, as well as an i-vector system. 
\item \textbf{Upper-bound comparison}: A large scale speaker verification experiment is done to set upper-bounds on performance by in-domain supervised methods. We present this to observe the margin of improvement of the proposed out-of-domain unsupervised method via additional higher level integration methods or layers. We note that our method only integrates 1~second level information while the i-vector and x-vector upper-bound methods use all the available data.
\end{enumerate}
  
The experimental setting for NPC training and the above experiments is described below.
\subsection{NPC Training Datasets}
\label{sec:Datasets}
Table \ref{tab:table_datasets} shows the training datasets along with their approximate total durations and number of contrastive samples created from each dataset. We train three different models individually on these datasets, and we call every trained model by the name of the dataset used for training along with the {NPC} prefix (for example, the NPC model trained on YoUSCTube data will be called as {NPC YoUSCTube}). 
\subsubsection{Tedlium dataset}
The Tedlium dataset is built from the Tedlium training corpora~\cite{rousseau2012ted}. It originally contained 666 unique speakers, but we have removed the 19 speakers which are also present in the Tedlium development and test sets (since the Tedlium dataset was originally developed for speech recognition purposes, it has speaker overlap between train and dev/test sets). 
The contrastive samples created from the Tedlium dataset are less noisy (compared to the case for YoUSCTube data as will be discussed next), because most of the audio streams in the Tedlium data are from a single speaker talking in the same environment for long (although there is some noise, for example, speech of the audience, clapping \textit{etc.}).

The reason for employing this dataset is two-fold:
First, the model trained on the Tedlium data will provide a comparison with the models trained on the Tedlium-Mix and YoUSCTube datasets for a validation of the short-term speaker stationarity hypothesis.
Second, since the test set of the speaker identification experiment will be from the Tedlium test data, this will help demonstrate the difference in performance for in-domain and out-of-domain evaluation.

\subsubsection{Tedlium-Mix dataset}
The Tedlium-Mix dataset is created mainly to validate the short-term speaker stationarity hypothesis (please see Section~\ref{sec:Validation of the short-term speaker stationarity hypothesis}). 
We create the Tedlium-Mix dataset by creating artificial dialogs through randomly concatenating utterances. 
Tedlium is annotated, so we know the utterance boundaries. We thus simulate a dialog that has a random speaker every other utterance of the main speaker. For every audio stream, we reject half of the total utterances, and between every two utterances we concatenate a randomly chosen utterance from a randomly chosen speaker (i.e. $S$, $R_1$, $S$, $R_2$, $S$, $R_3, \dots$  where $S$'s are the utterances of the main speaker and $R_i$ (for $i=1,2,3,\dots$) is a random utterance from a randomly chosen speaker i.e. a random utterance from another Ted recording). In this way we create the Tedlium-Mix dataset having a speaker change after every utterance for every audio stream. It also has almost the same size as the Tedlium dataset.

\subsubsection{YoUSCTube dataset}
\label{subsubsec:YoUSCTube dataset}
A large amount of various types of audio data has been collected from YouTube to create the YoUSCTube dataset. We have chosen YouTube for this purpose because of virtually unlimited supply of data from diverse environments. The dataset has multilingual data including English, Spanish, Hindi and Telugu from heterogeneous acoustic conditions like monologues, multi-speaker conversations, movie clips with background music and noise, outdoor discussions \textit{etc}.

\subsubsection{Validation data}
\label{sub:validation}
The Tedlium development set (8 unique speakers) has been used as  validation data for all training cases.
We used the utterance start and end time stamps and the speaker IDs provided in the transcripts of the Tedlium dataset to create the validation set so that it does not have any noisy labels.

\subsection{Data for speaker identification experiment}
The Tedlium test set (11 unique speakers from 11 different Ted recordings) has been employed for the speaker identification experiment. 
Similar to the development dataset, it has start and end time of every utterance for every speaker as well as the speaker IDs. 
We have extracted the utterances from every speaker, and all utterances of a particular speaker have been assigned the corresponding speaker ID. 
Those have been used for creating the experimental scenarios for speaker classification (Section~\ref{sec:Frame-level speaker classification} and Section~\ref{sec:Utterance-level speaker classification}). 
Similar to the validation set, the labels of this dataset are very clean since they are created using the human-labeled speaker IDs.

\subsection{Data for speaker verification experiment}
A recently released large speaker verification corpus, VoxCeleb (version 1)~\cite{nagrani2017voxceleb} is employed for the speaker verification experiment.
It has a total of 1251 unique speakers with $\sim 154K$ utterances at 16KHz sample rate.
The average number of sessions per speaker is 18.
We use the default development and test split provided with the dataset and mentioned in~\cite{nagrani2017voxceleb}.

\subsection{Feature and model parameters}
\label{sec:Feature extraction}
We employ 40 dimensional MFCC features computed from 16KHz audio with 25\textit{ms} window and 10\textit{ms} shift using the Kaldi toolkit~\cite{Povey_ASRU2011}. 
We  choose $d=100$ frames (1\textit{s}), and $\Delta=200$ frames (2$s$). 
Therefore, each window is a $100 \times 40$ matrix, and we feed this to the first CNN layer of our network (Fig.~\ref{fig:DNN}).
The employed model has a total 1.8M parameters and it has been trained using RMSProp optimizer~\cite{tieleman2012lecture} with a learning rate of $10^{-4}$ and a weight decay of $10^{-6}$.
The held out validation set (Section~\ref{sub:validation}) has been used for model selection.

\begin{figure}[t]
	\centering
	\includegraphics[trim=100 230 100 250, clip,width=\columnwidth]{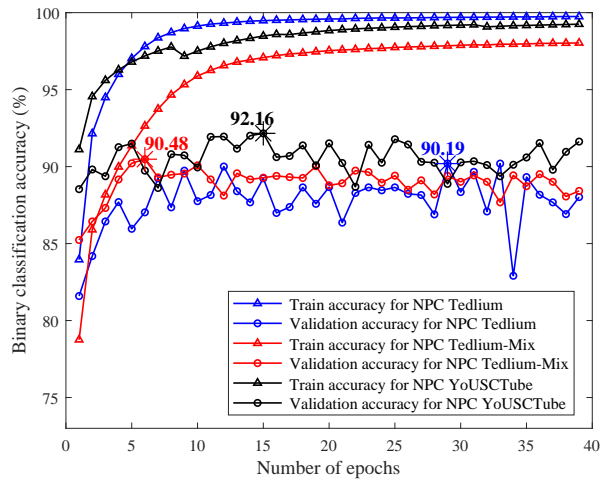}
	\caption{Binary classification accuracies of classifying genuine or impostor pairs for NPC models trained on the Tedlium, Tedlium-Mix, and YoUSCTube datasets. Both training and validation accuracies are shown. The best validation accuracies for all the models are marked by big stars (*).}
	\label{fig:STSShyp_graphs}
\end{figure}

\section{Experimental Results}
\label{sec:results}

\subsection{Convergence curves}
\label{sec:convergence curves}
Fig.~\ref{fig:STSShyp_graphs} shows the convergence curves in terms of binary classification accuracies of classifying genuine or impostor pairs for training the DNN model separately in different datasets along with the corresponding validation accuracies. The development  set for calculating the validation accuracy is same for all the training sets and it doesn't contain any noisy samples. In contrast, our training set is  noisy since it's unsupervised and based on the short-term stationarity in assigning same/different class speaker pairs.

We can see from Fig.~\ref{fig:STSShyp_graphs} that {NPC Tedlium} reaches almost $100\%$ training accuracy, but {NPC Tedlium-Mix} converges at a lower training accuracy as expected. 
  This is due to the larger portion of noisy samples present in the Tedlium-Mix dataset that arise from the artificially introduced fast speaker changes and the simultaneous hypothesis of short-term speaker stationarity\footnote{The corpus is created by mixing turns. This means that there are $54,778$ speaker change points in the $115$ hours of audio.  However in this case we assumed that there are no speaker changes in consecutive frames. If the change points were uniformly distributed then that would result in an upper-bound of 87\%.}.
However this doesn't hurt the validation accuracy on the development set, which is both distinct from training set and correctly labeled:  we obtain $90.19\%$ and $90.48\%$ for {NPC Tedlium} and {NPC Tedlium-Mix} trained-models respectively. We believe this is because the model is correctly learning to \emph{not} label some of the assumed same-speaker pairs as same-speaker when there is a speaker change that we introduced via our mixing, due to the large amounts of correct data that compensate for the smaller-amount of mislabeled pairs.

The {NPC YoUSCTube} model reaches much better training accuracy than the {NPC Tedlium-Mix} model even with fewer epochs. 
This points to both increased robustness due to the increased data variability and also that speaker-changes in real dialogs are not as fast as we simulated in the Tedlium-Mix dataset.
It is interesting to see that the {NPC YoUSCTube} model achieves a little better validation accuracy ($92.16\%$) than the other two models even when the training dataset had no explicit domain overlap with the validation data. We think, this is because of the huge size (approximately 6 times larger in size than the Tedlium dataset) and widely varying types of acoustic environments of the YoUSCTube dataset. 

\subsection{Validation of the short-term speaker stationarity hypothesis}
\label{sec:Validation of the short-term speaker stationarity hypothesis}
Here we analyze the validation accuracies obtained by the NPC models trained separately on the Tedlium and Tedlium-Mix datasets. From Fig.~\ref{fig:STSShyp_graphs} it is quite clear that both models could achieve similar validation accuracies, although the Tedlium-Mix dataset has audio streams containing speaker changes at every utterance and the Tedlium dataset contains mostly single-speaker audio streams. The reason is that even though there are frequent speaker turns in the Tedlium-Mix dataset, the short length of context ($d=100$ frames $=$ 1\textit{s}) chosen to learn the speaker characteristics ensures that the total number of correct same-speaker pairs dominate the falsely-labeled same-speaker pairs. Therefore the sudden speaker changes are of little impact and do not deteriorate the performance of neural network on the development set. 
This result  validates the short-term speaker stationarity hypothesis.

\begin{figure*}
	\centering
	\begin{subfigure}[b]{0.49\textwidth}
		\includegraphics[width=\textwidth]{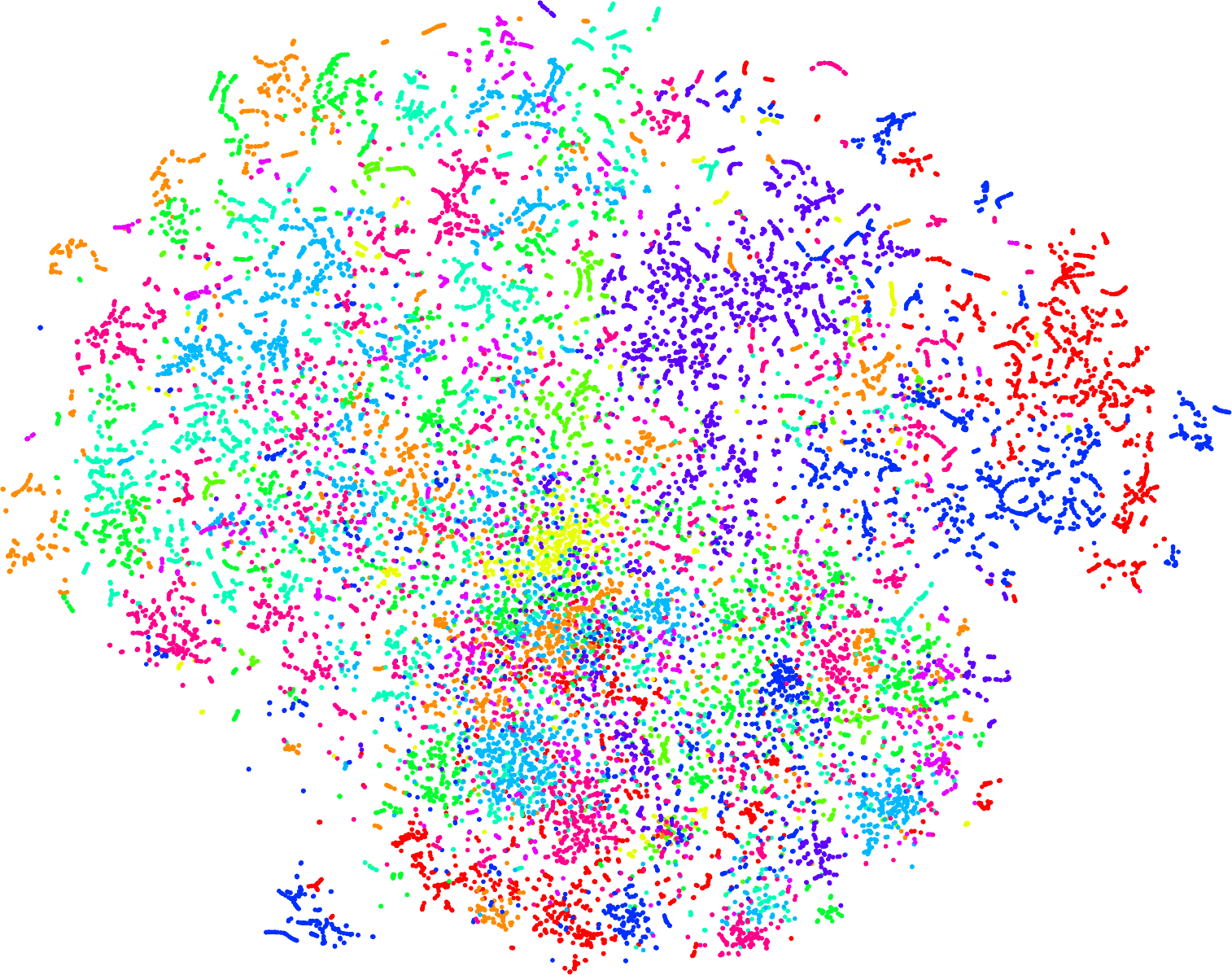}
		\caption{{MFCC}}
		\label{fig:vis_mfcc}
	\end{subfigure}
	\begin{subfigure}[b]{0.49\textwidth}
		\includegraphics[width=\textwidth]{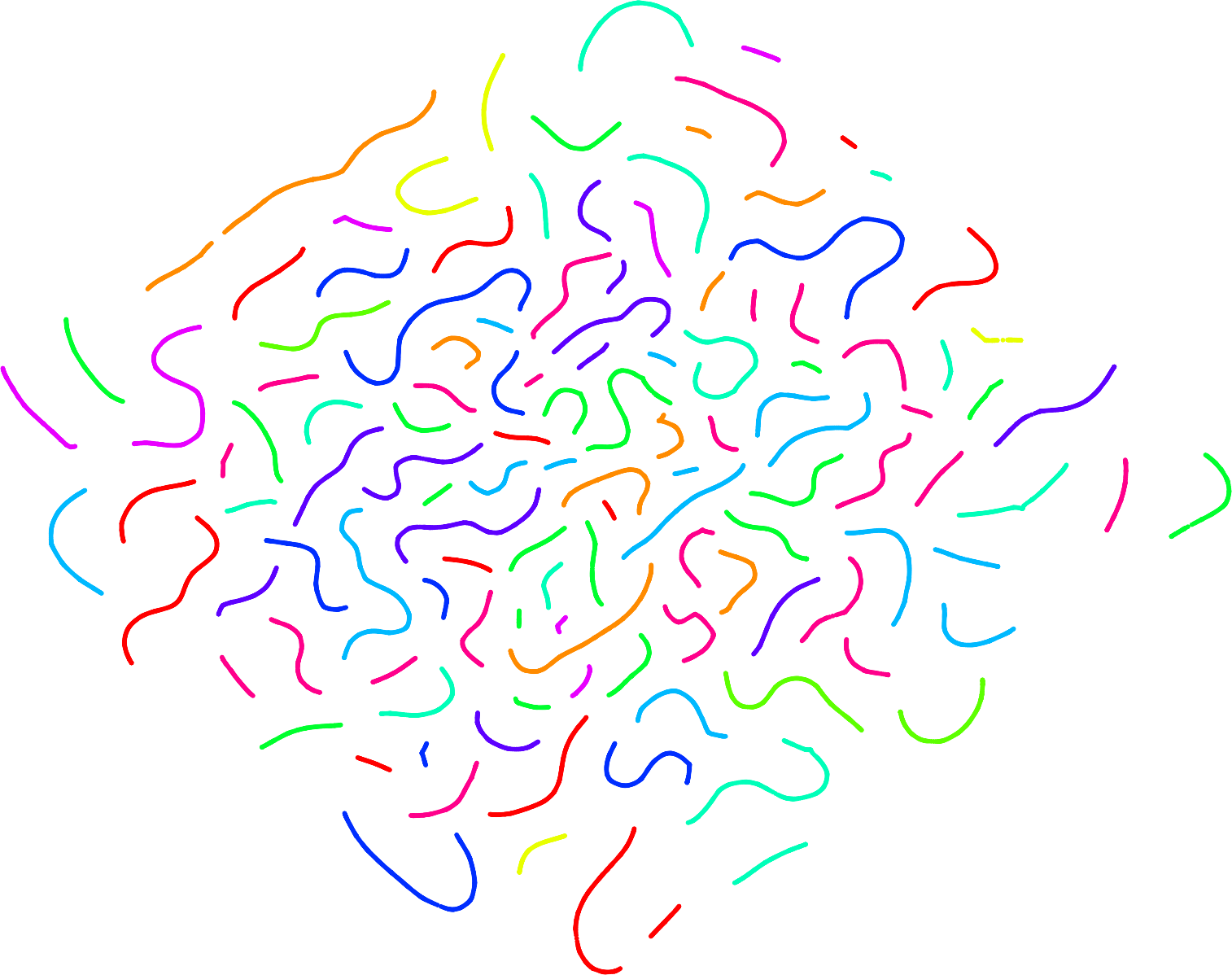}
		\caption{{MFCC stats}}
		\label{fig:vis_mfcc_stats}
	\end{subfigure}
	\begin{subfigure}[b]{0.49\textwidth}
		\includegraphics[width=\textwidth]{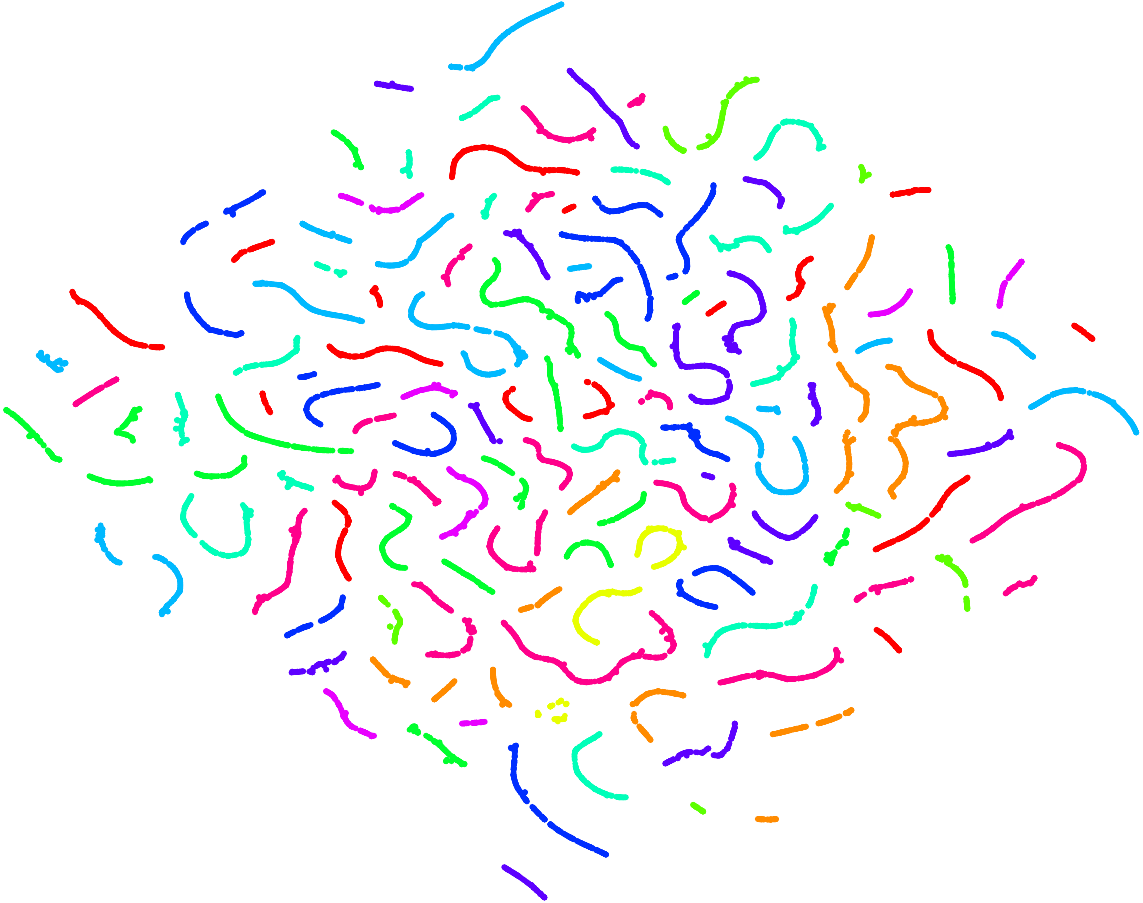}
		\caption{{i-vector}}
		\label{fig:vis_ivectors}
	\end{subfigure}
	\begin{subfigure}[b]{0.49\textwidth}
		\includegraphics[ width=\textwidth]{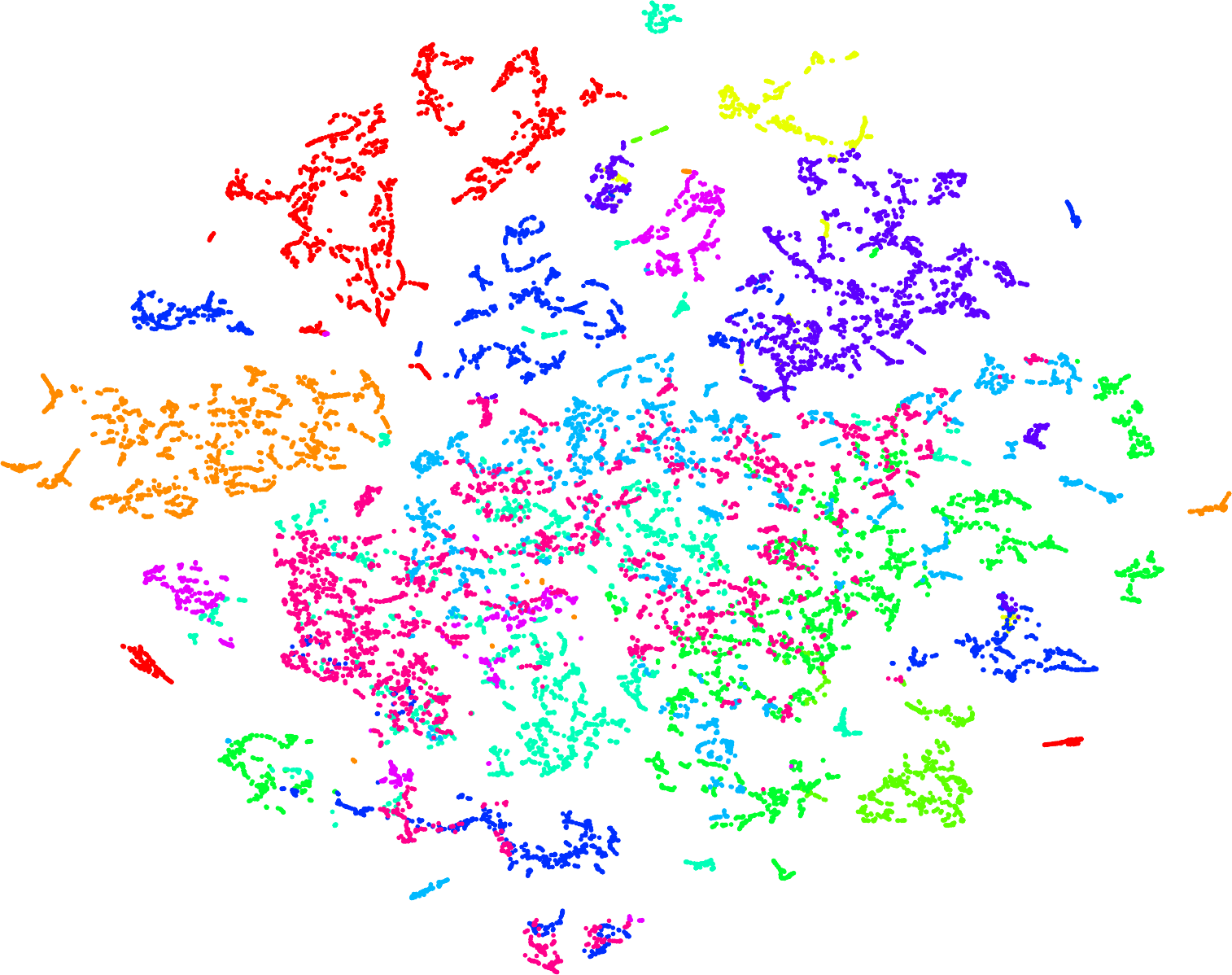}
		\caption{{NPC YoUSCTube Cosine}}
		\label{fig:vis_npc_cosine}
	\end{subfigure}
	\begin{subfigure}[b]{0.49\textwidth}
		\includegraphics[ width=\textwidth]{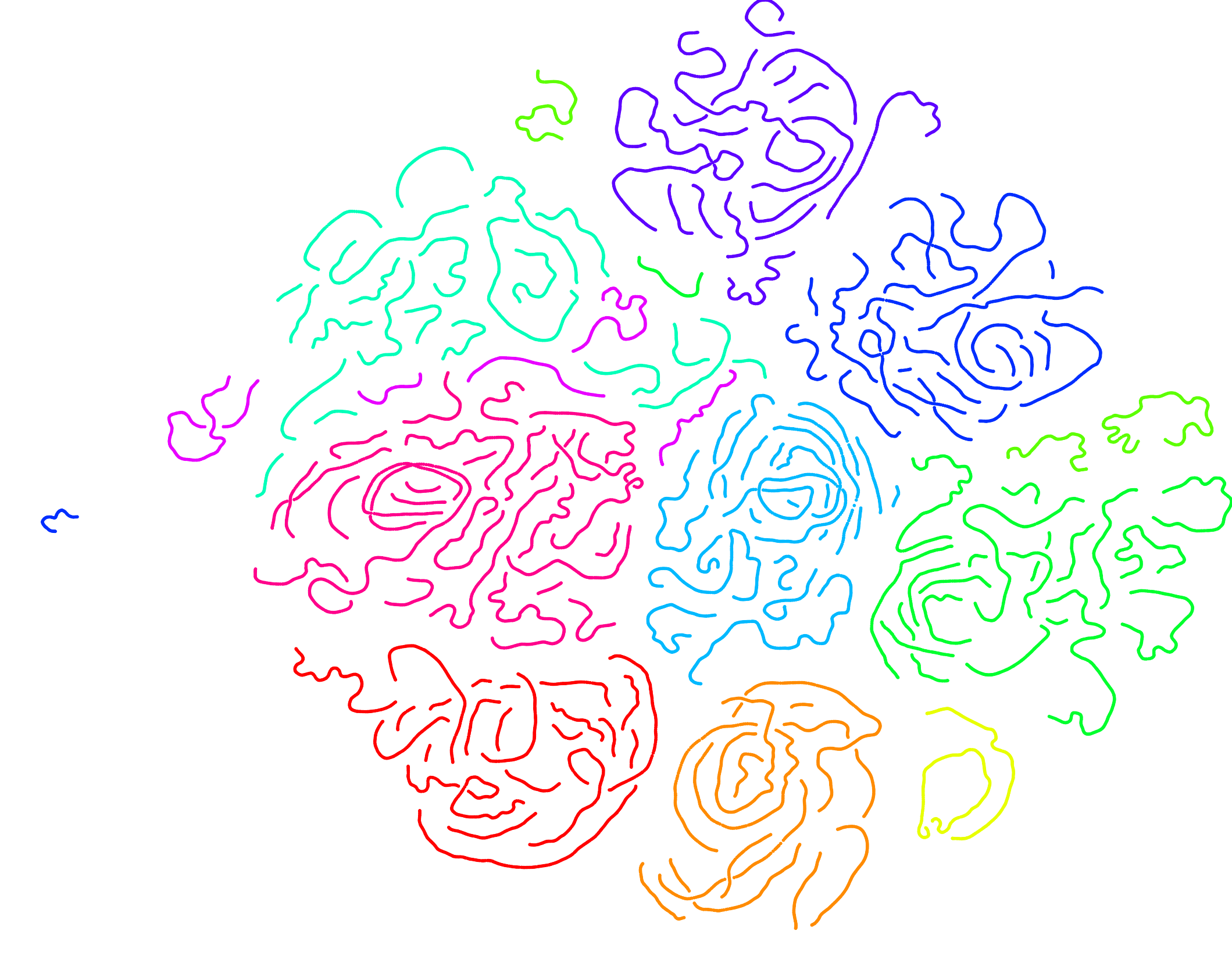}
		\caption{{NPC YoUSCTube Cross Entropy}}
		\label{fig:vis_npc_youtube}
	\end{subfigure}
	\begin{subfigure}[b]{0.49\textwidth}
		\includegraphics[width=\textwidth]{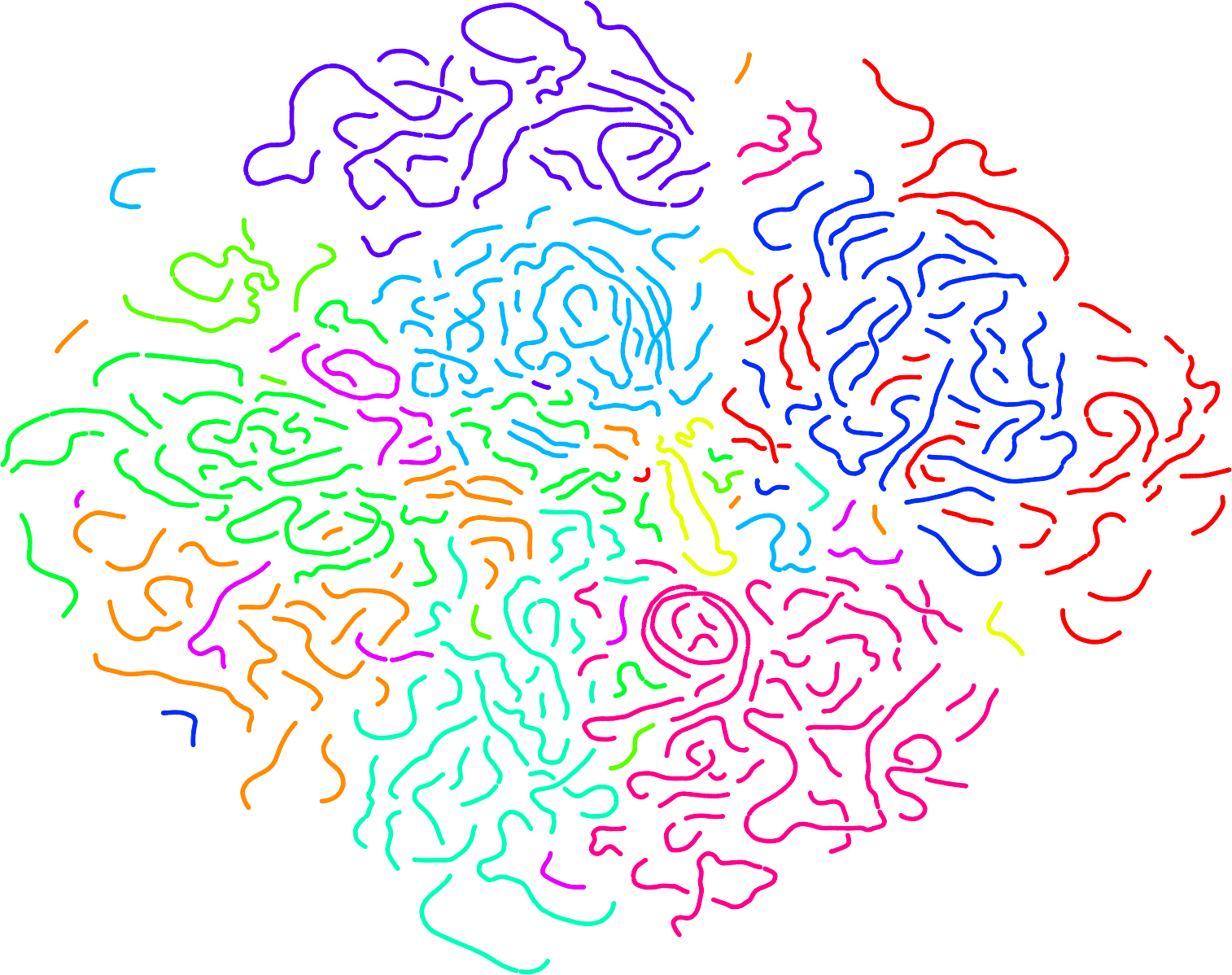}
		\caption{{NPC Tedlium Cross Entropy}}
		\label{fig:vis_npc_ted}
	\end{subfigure}
	\caption{t-SNE visualizations of the frames of different features for the Tedlium test data containing 11 speakers (2 utterances per speaker). Different colors represent different speakers.%
	}\label{fig:visualization}
\end{figure*}

\subsection{Experiments: \textbf{Speaker identification evaluation}}
\label{sec:Experiments Speaker identification evaluation}

\subsubsection{Frame-level Embedding visualization}
\label{sec:Embedding visualization}
Visualization of high dimensional data is vital in many scenarios because it can reveal the inherent structure of the data points. For speaker characteristics learning, visualizing the employed features can manifest the clusters formed around different speakers and thus demonstrate the efficacy of the features. We use t-SNE visualization~\cite{maaten2008visualizing} for this purpose. 
We compare the following features (the terms in \textbf{boldface} show the names we will use to call the features).
\begin{enumerate}

\item {\textbf{MFCC}}: Raw  MFCC features.

\item {\textbf{MFCC stats}}: This is generated by moving a sliding window of 1\textit{s} along the raw MFCC features with a shift of 1 frame (10\textit{ms}) and taking the statistics (mean and standard deviation in the window) to produce a new feature stream. This is done for a fair comparison of MFCC and the embeddings (since the embeddings are generated using 1\textit{s} context).

\item {\textbf{NPC YoUSCTube Cross Entropy}}: Embeddings extracted with {NPC YoUSCTube} model using cross entropy loss. 

\item {\textbf{NPC Tedlium Cross Entropy}}: Embeddings extracted with {NPC Tedlium} model using cross entropy loss.

\item {\textbf{NPC YoUSCTube Cosine}}: Embeddings extracted with {NPC YoUSCTube} model using cosine embedding loss.  

\item {\textbf{i-vector}}: 400 dimensional i-vectors extracted independently every 1\textit{s} using a sliding window with 10\textit{ms} shift. 
  The i-vector system (Kaldi VoxCeleb v1 recipe) was trained on the VoxCeleb dataset~\cite{nagrani2017voxceleb} (16~KHz audio). It is not possible to train an i-vector system on YoUSCTube since it contains no labels on speaker-homogeneous regions.

\end{enumerate}

Fig.~\ref{fig:visualization} shows the 2 dimensional t-SNE visualizations of the frames (at 10\textit{ms} resolution) of the above features extracted from the Tedlium test dataset containing 11 unique speakers. 
For better visualization of the data, we chose only 2 utterances from every speaker, and the feature frames from a total of 22 utterances become our input dataset for the t-SNE algorithm. 
From Fig.~\ref{fig:visualization} we can see that the raw MFCC features are very noisy, but the inherent smoothing applied to compute {MFCC stats} features help the features of the same speaker to come closer. However we notice that although some same-speaker features cluster in  lines, these lines are far apart in the space, which denotes that the MFCC features capture additional information. For example we see that the speaker denoted with Green occupies both the very left and very right parts of the t-SNE space.

The i-vector plot looks similar to the MFCC stats and does not cluster the speakers very well.
This is consistent with existing literature~\cite{kanagasundaram2011vector} that showed that i-vectors do not perform well for short utterances especially when the training utterances are comparatively longer. 
In Section~\ref{sec:Utterance-level speaker classification}, we will see that the utterance-level i-vectors perform much better for speaker classification.

The {NPC YoUSCTube Cosine} embeddings underperform the cross entropy-based methods possibly because of poorer convergence as we observed during training. 
They are also noisier than MFCC stats and i-vectors, indicating that even a little change in the input (just 10\textit{ms} of extra audio) perturbs the embedding space, which might not be desirable.

The {NPC YoUSCTube Cross Entropy} and {NPC Tedlium Cross Entropy}  embeddings provide much better distinction between different speaker clusters. Moreover, they also provide much better cluster compactness compared to the MFCC and i-vector features. 

Among the NPC embeddings, {NPC YoUSCTube Cross Entropy} features provide possibly the best tSNE visualization. They even produce better clusters than {NPC Tedlium Cross Entropy}, although the latter one is trained on in-domain data. The larger size of YoUSCTube dataset might be the reason behind this observation.

\subsubsection{Frame-level speaker identification}
\label{sec:Frame-level speaker classification}
We perform frame-level speaker identification experiments on the Tedlium development set (8 speakers) and the Tedlium test set (11 speakers). 
By frame-level classification we mean that every frame in the utterance is independently classified as to its speaker ID.

The reason for evaluating with frame-level speaker classification is that better frame-level performance conveys the inherent strength of the system to derive short-term features that carry speaker-specific characteristics. 
It also shows the possibility to replace MFCCs with the proposed embeddings by incorporating in systems such as~\cite{yamada2013improvement} and ~\cite{ghalehjegh2015deep}.

\begin{table}[!t]
	\caption{Frame-level speaker classification accuracies of different features with kNN classifier (k=1). All features below are trained on unlabeled data except i-vector which requires speaker-homogeneous files.}
	\label{tab:speaker_identification}
	\centering
  \includegraphics[width=\columnwidth]{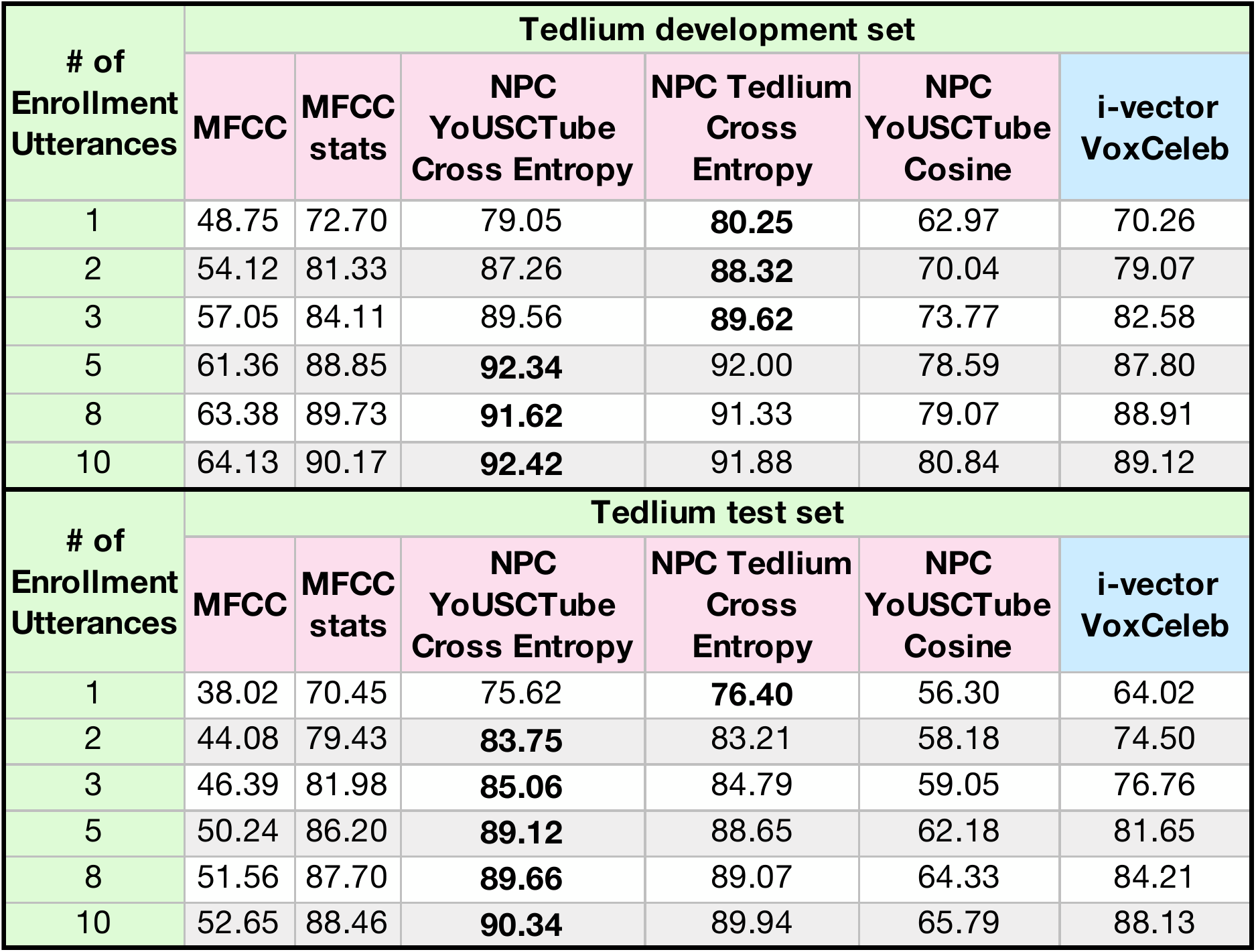}\\
\end{table}

Table~\ref{tab:speaker_identification} shows a detailed comparison between the 6 different features described in Section~\ref{sec:Embedding visualization} in terms of frame-wise speaker classification accuracies.
We have tabulated the accuracies for different number of enrollment utterances (in other words training utterances for the speaker ID classifier) per speaker.
We have used kNN classifier (with k=1) for speaker classification. 
The reason for using such a naive classifier is to reveal true potential of the features, and not to harness strength of the classifier. 
We have repeated each experiment 5 times and the average accuracies have been reported here. 
Each time we have held out 5 random utterances from each speaker for testing. 
The same seen (enrollment) and test utterances have been used for all types of features and in all cases the test and enrollment  sets are distinct. 

From Table~\ref{tab:speaker_identification} we can see that {MFCC stats} perform much better than raw {MFCC} features. We think the reason is that the raw features are much noisier than the {MFCC stats} features because of the implicit smoothing performed during the statistics computation. 
The {NPC YoUSCTube} and {NPC Tedlium} models with cross entropy loss perform pretty similarly (for test data, {NPC YoUSCTube} even performs better) even though the former one is trained on out-of-domain data. 
This highlights the benefits and possibilities of employing out-of-domain unsupervised learning using publicly available data. 
{NPC YoUSCTube Cosine} doesn't perform well compared to other NPC embeddings. 
The i-vectors perform worse than NPC embeddings and MFCC stats for frame-level classification due to the reasons discussed in Section~\ref{sec:Embedding visualization} and as reported by \cite{kanagasundaram2011vector}.

\begin{figure*}
	\centering
	\begin{subfigure}[b]{0.49\textwidth}
		\includegraphics[trim=75 270 90 280 , clip, width=\textwidth]{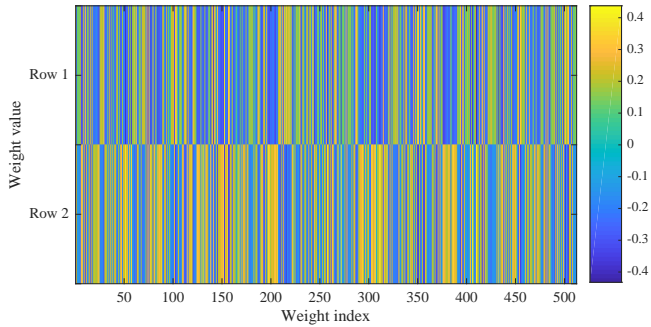}
		\caption{Two rows of the full $512\times 2$ matrix shown as an image.}
		\label{fig:W_image}
	\end{subfigure}
	~ %
	\begin{subfigure}[b]{0.49\textwidth}
		\includegraphics[trim=90 270 90 270, clip,width=\textwidth]{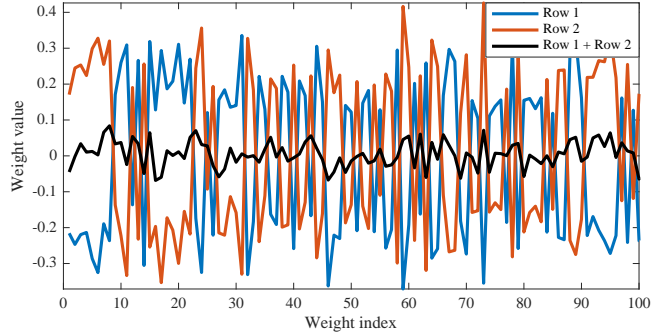}
		\caption{Weights in the two rows plotted, along with their sum.}
		\label{fig:W_zoomed}
	\end{subfigure}
	\caption{Visualization of the learned weights in the last fully connected (FC) layer: (a) The matrix of the last FC layer. Note the opposite signs of the weight values in `Row 1' and `Row 2' for a particular weight index. (b) This figure shows a zoomed version of the figure in (a) for only 100 weights. Note the mirrored nature of the weights values in `Row 1' and `Row 2', and oscillation of their sum around zero.}
	\label{fig:weights_visualization}
\end{figure*}
\subsubsection{Analyzing network weights} 
\label{sec:Analyzing network weights}
We have seen in the previous experiments that the embeddings learned using the cross entropy loss performed better than those learned through minimizing the cosine loss.
Here we analyze the learned weights in the last fully connected (FC) layer (size = $512\times 2$) in the network that uses cross entropy loss.
From Fig.~\ref{fig:weights_visualization} we can see that the weights are learned in a way such that the weight value for a particular position of the first embedding, $w_{1,k}$, is approximately of same value and opposite sign of the weight value for that particular position in the second embedding, $w_{2,k}$ (please see Equation~\ref{eqn:g(x1,x2)} for the notations).
Experimentally, for {NPC YoUSCTube Cross Entropy} model, we found the mean of absolute value of $w_{1,k}+w_{2,k}$ to be $0.0284$, with a standard deviation of $0.0206$ (mean and standard deviations computed over all the embedding dimensions, \textit{i.e.} $k$ varying from $1$ to $D=512$).  
The two bias values we found are $1.0392$ and $-0.9876$.
In other words, the experimental evidence shows 
\begin{align*}
	w_{1,k} \approx -w_{2,k} \\
	\text{and, }
	b_1 \approx -b_2.
\end{align*}
 One possible and intuitive explanation would be that the individual absolute weight values provide importance to different dimensions/features in the embedding (this has also been explained in~\cite{koch2015siamese}). The mirrored nature of weights and biases are possibly ensuring cancellation of same-speaker embeddings while ensuring maximization of impostor pair distance. 
 For example, if $w_{1,k}$ is a high positive number then it ensures higher contribution of the $k^{th}$ dimension of $\mathbf{L(x_1,x_2)}$ in the softmax output for the genuine pair class (since, $w_{1,k}|f(\mathbf{x_1})_k-f(\mathbf{x_2})_k|$ will be higher).
 On the other hand, $w_{2,k} \approx -w_{1,k}$ is ensuring ``equally lower" contribution of $|f(\mathbf{x_1})_k-f(\mathbf{x_2})_k|$ to the probability of the input to be an impostor pair.
   
For the cosine embedding loss, these automatically learned importance weights are not present, which might be the reason for under performing the cross entropy embeddings; all embedding dimensions are equally contributing to the loss. 

\begin{table}[!ht]
	\caption{Utterance-level speaker classification accuracies of different features with kNN classifier (k=1). Red italics indicates the best performing single feature classification result while bold text indicates the best overall performance.}
	\label{tab:speaker_identification_uttr}
	\centering
	\includegraphics[width=\columnwidth]{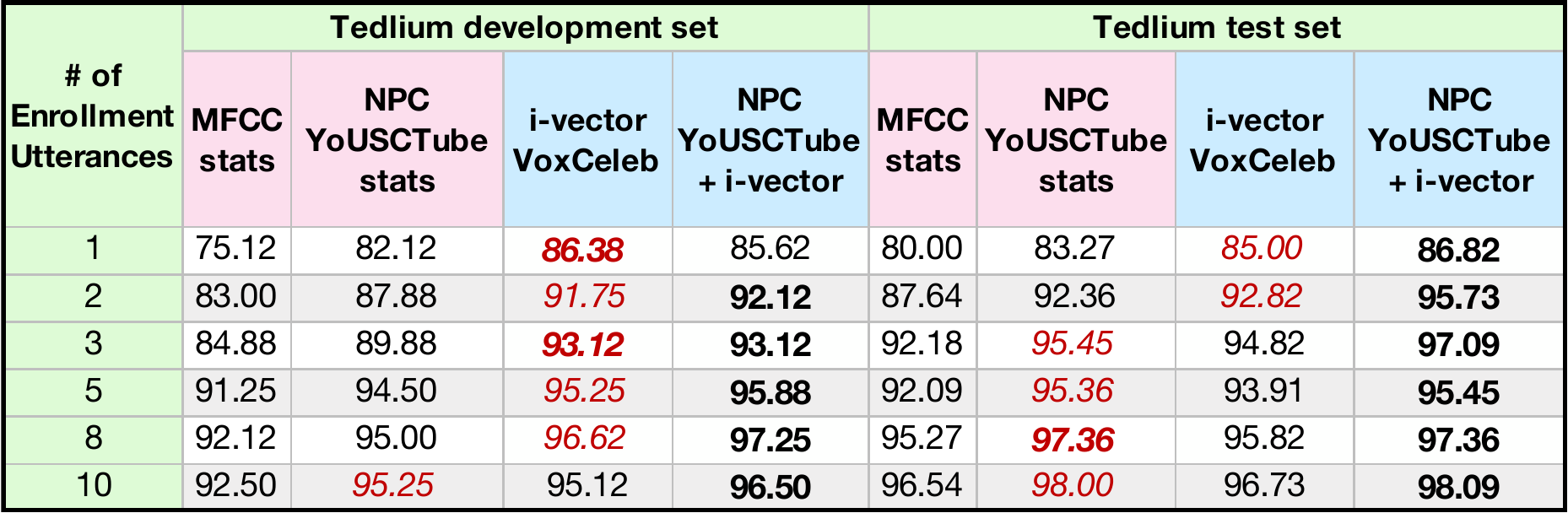}
\end{table}

\subsubsection{Utterance-level speaker identification}
\label{sec:Utterance-level speaker classification}
Here we are interested in  utterance-level speaker identification task. 
We compare {NPC YoUSCTube Cross Entropy} (out-of-domain (OOD) YouTube), {MFCC}, and {i-vector} (OOD VoxCeleb) methods.
For MFCC and NPC embeddings, we  calculate the mean and standard deviation vectors over all frames in a particular utterance, and concatenate them to produce a single vector for every utterance. 
For i-vector, we  calculate one i-vector for the whole utterance using the same i-vector system as mentioned in Section~\ref{sec:Embedding visualization}.

We applied LDA (trained on development part of VoxCeleb) to project the 400 dimensional i-vectors to a 200 dimensional space.
This gave better performance for i-vectors (also observed in literature~\cite{snyder2017deep}) and let us compare unsupervised NPC embeddings with the best possible i-vector configuration.
We again classify using k-NN classifier with $k=1$, as explained  in Section~\ref{sec:Frame-level speaker classification} to focus on the feature performance and not on the next-layer of trained classifiers.

Table~\ref{tab:speaker_identification_uttr} shows the classification accuracies for different features with increasing number of enrollment utterances
\footnote{Note that due to the small-size window for our feature, even two  utterances provide significant information; hence we do not see significant change as the enrollment utterances increase.}.
In each enrollment scenario, 5 randomly held-out utterances from each of the 11 speakers have been used for testing, and the process has been repeated 20 times to report the average accuracies.
Both i-vectors and {NPC YoUSCTube} embeddings perform similarly.
It is interesting to note the complementarity of the concatenated i-vector-embedding feature. 
From Table~\ref{tab:speaker_identification_uttr} we can see that the {NPC YoUSCTube Cross Entropy} + {i-vector} perform the best for almost all the cases.  

An additional important point is that the  classifier used is the simple 1-Nearest Neighbor classifier. 
So, we believe that the highly non-Gaussian nature of the embeddings (as can be observed from Fig.~\ref{fig:visualization}) might \textit{not} be captured well by the 1-NN since it is based on Euclidean distance which will under perform in complex manifolds as we observe with NPC embeddings.
This motivates future work in higher-layer, utterance-based, neural network-derived features that build on top of these embeddings.     

\begin{table}[t]
	\caption{Speaker verification on VoxCeleb v1 data. i-Vector and x-Vector use the full utterance in a supervised manner for evaluation while the proposed embedding operates at the 1~second window with a simple statistics (mean+std) over an utterance.}
	\label{tab:vox_SV}
	\centering
	\includegraphics[width=\columnwidth]{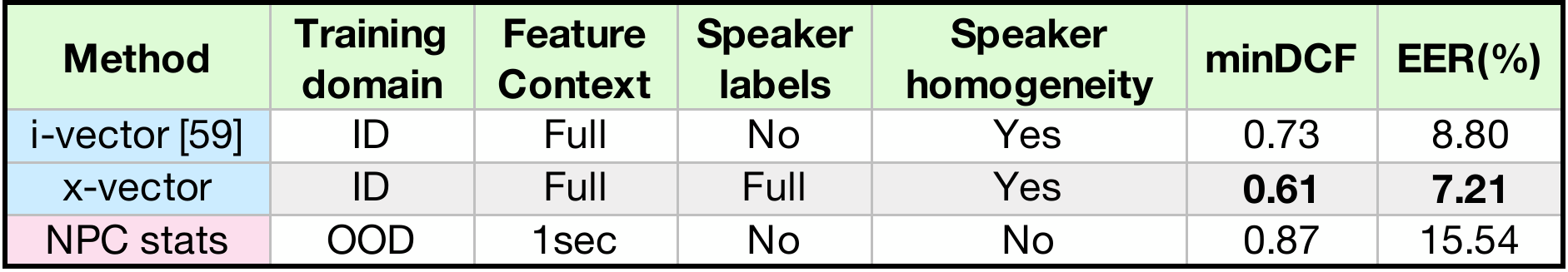}
\end{table}
\subsection{Experiments: \textbf{Upper-bound comparison}}
\label{sec:UpperBound}
Table~\ref{tab:vox_SV} compares performance of i-vector, x-vector~\cite{snyder2018x}, and the proposed NPC embeddings for the speaker verification task on Voxceleb v1 data using the default Dev and Test splits~\cite{nagrani2017voxceleb} distributed with the dataset.

We want to highlight that since the assumption for our system is that we have absolutely no labels during DNN training (in fact our YouTube downloaded data are not even guaranteed to be speech!), the comparison with x-vector or i-vector is highly asymmetric.
To simplify this explanation:
\begin{itemize}
	\item Our proposed method uses ``some random audio'': completely unsupervised and challenging data.
	\item i-vector uses ``speech'' with labels on ``speaker homogeneous regions'': unsupervised with a supervised step on clean data.
	\item x-vector uses ``speech'' with labels of ``id of speaker'': completely supervised on clean data.
\end{itemize}
Moreover, i-vector and x-vector here are trained on in-domain (ID) Dev part of the VoxCeleb dataset. 
On the other hand, the NPC model is trained out-of-domain (OOD) on unlabeled YouTube data.
Please note that here ``out-of-domain'' refers to the generic characteristics of the YoUSCTube dataset compared to the Voxceleb dataset. For example, the Voxceleb dataset was mined using the keyword ``interview''~\cite{nagrani2017voxceleb} along with the speaker name, and the active speaker detection~\cite{nagrani2017voxceleb} ensured active presence of that speaker in the video. On the other hand, the YoUSCTube dataset is mined without any constraints thus generalizing more to realistic acoustic conditions~(see ~\ref{subsubsec:YoUSCTube dataset}). Moreover, having only celebrities~\cite{nagrani2017voxceleb} in the Voxceleb dataset helped it to find multiple sessions of the same speaker, which subsequently helped the supervised DNN models to be more channel-invariant. However, such freedom is not available in the YoUSCTube dataset, thus paving a way to build unsupervised models that can be trained or adapted in such conditions.

Finally, the features employed by i-vector and x-vector employ the whole utterance of average length 8.2\textit{s} (min=4\textit{s}, max=145\textit{s})~\cite{nagrani2017voxceleb} while the NPC model is only producing 1~second estimates. While we do intend to incorporate more contextual learning for longer sequences, in this work we are focusing on the low-level feature and hence employ statistics (mean and std) of the embeddings. This is suboptimal and creates an uninformed information bottleneck, however it is a necessary and easy way to establish an utterance-based feature, thus enabling comparison with the existing methods.

For all the above reasons we expect that any evaluation with i-vector and x-vector can only be seen as a very upper-bound and we don’t expect to beat either of these two in performance.

The i-vector performance is as reported in~\cite{nagrani2017voxceleb}.
No data augmentation is performed for x-vector for a fair comparison.

To maintain standard scoring mechanisms we employed LDA to project the embeddings on a lower dimensional space and, then PLDA scoring as in \cite{snyder2018x,nagrani2017voxceleb}. 
The same VoxCeleb Dev data is utilized to train LDA and PLDA models for all methods for a fair comparison.
The LDA dimension is 200 for x-vector and i-vector~\cite{nagrani2017voxceleb} systems,  and 100 for NPC system.
We report the minimum normalized detection cost function (minDCF) for $P_{target}=0.01$ and Equal Error-Rate (EER).
We can see that the best in-domain supervised method is 30\% better than unsupervised NPC in terms of minDCF.

\section{Discussion and Future Work}
\label{sec:Discussion_FutureWork}

\subsection{Discussion}
Based on the visualization of Fig.~\ref{fig:visualization} and the experiments of Section~\ref{sec:Experiments Speaker identification evaluation} we have established that the resulting embedding is capturing significant information about the speakers' identity. The feature has shown to be quite better than using knowledge driven features such as MFCCs or statistics of MFCCs and even more robust than supervised features such as i-vector operating on 1~second windows. Importantly the proposed embedding showed extreme portability by operating better on the Tedlium dataset when trained on larger amounts of random audio from the collected YoUSCTube corpus than when trained in-domain on the Tedlium dataset itself.

Also importantly we have shown in sections \ref{sec:convergence curves} that if we on purpose create a fast changing dialog by mixing the Tedlium utterances, the short-term stationarity hypotheses still holds. This encourages the use of unlabeled data.

Evaluating this embedding however is challenging as its use is not obvious until it is used for a full blown speaker identification framework. This requires several more stages of development that we will discuss further in this work, along with discussing the shortcomings of this embedding. However, we can, and we are, providing some early evidence that the embedding does indeed capture significant information about the speaker.

In Section \ref{sec:Utterance-level speaker classification} we present results that compare an utterance-based classification system on the Tedlium data. We are comparing the i-vector system optimized for utterance-level classification, and which employs supervised data, with a very simple statistic (mean and std) of our proposed unsupervised embedding. We show that our embedding provides very robust results that are comparative to the i-vector system. 
The shortcoming of this comparison, is that the utterances are drawn from the Tedlium dataset, and they are likely also incorporating channel information. 
We provide some suggestions in overcoming this shortcoming further in this section.

We proceeded, in Section \ref{sec:UpperBound}, to present results that compare an utterance-based classification system on the VoxCeleb v1 test. Here we wanted to provide an upper-bound comparison. We evaluated i-vector and x-vector VoxCeleb trained supervised methods. These methods are able to employ the full utterance as a single observation, while the proposed embedding only operates on a $<$~1~second resolution, hence we again aggregate via an uninformed information bottleneck (mean and std). We see that despite the information bottleneck and complete unsupervised and out of domain nature of the experiment our proposed system still achieves an acceptable performance with a 30\% worse minDCF than x-vector.

\subsection{Future work}

The above observations and analysis provide many directions for future work.

Given that all our same-speaker examples come from the same channel, we believe that the proposed embedding captures both channel and speaker characteristics. This provides an opportunity for data augmentation, and hence reduction of the channel influence. In future work we intend to augment the near-by frames such that contextual pairs are coming from a range of different channels through augmentation.

This also provides another opportunity for joint channel and speaker learning. Through the above augmentation we can jointly learn same vs different speakers and same vs different channels, thus providing disentanglement and more robust speaker representations.

Further, triplet learning~\cite{li2017deep}, especially with hard triplet mining, has been shown to provide improved performance and we intend to use such an architecture in future work to directly optimize intra- and inter-class distances in the manifold.

One additional opportunity for improvement is to employ a larger neural network.  We employed a CNN with only 1.8M parameters for our training (as an initial try to check the validity of the proposed method). But, recent CNN-based speaker verification systems employ much deeper networks (\eg VoxCeleb's baseline CNN comprises of 67M parameters). We think utilizing recent state-of-the-art deep architectures will improve performance of the proposed technique for large scale speaker verification experiments.

Finally, and more applicable to the speaker ID task, we need embeddings that can capture information from longer sequences. As we see in Section~\ref{sec:UpperBound} the supervised speaker identification methods are able to exploit longer term context while the proposed embedding is only able to serve as a short-term feature. This requires either supervised methods, towards higher level information integration, or more in alignment with our interests of better unsupervised context exploitation. For example we can employ a better aggregation mechanism via unsupervised diarization using this embedding to identify speaker-homogeneous regions and then employ Recurrent Neural Networks (RNN)~\cite{Goodfellow-et-al-2016} towards longitudinal information integration.

\section{Conclusion}
\label{sec:conclusion}

In this paper, we proposed an unsupervised technique to learn speaker-specific characteristics from unlabeled data that contain any kind of audio, including speech, environmental sounds, and multiple overlapping speakers of many languages.

The proposed system exploits the short-term active-speaker stationarity hypothesis to create contrastive samples from unlabeled data, and feed them into a deep convolutional siamese network which learns the NPC embeddings by learning to classify same vs different speaker pairs.

We trained the proposed siamese model on both the YoUSCTube and Tedlium training sets.
We performed two sets of evaluation experiments: a closed set speaker identification experiment, and a large scale speaker verification experiment for upper-bound comparison.
The NPC embeddings outperform i-vectors at frame-level speaker identification, and provide complementary information to i-vectors at the utterance-level speaker identification task.

As an upper-bound task we employed the VoxCeleb speaker verification set. As expected NPC embeddings underperform in-domain supervised x-vector and in-domain i-vector methods.

The analysis of the proposed out-of-domain unsupervised method with the in-domain supervised methods helps identify challenges and raises a range of opportunities for future work, including in longitudinal information integration and in introducing robustness to channel characteristics.

\section*{Acknowledgment}
The U.S. Army Medical Research Acquisition Activity, 820 Chandler Street, Fort Detrick MD 21702- 5014 is the awarding and administering acquisition office. This work was supported by the Office of the Assistant Secretary of Defense for Health Affairs through the Military Suicide Research Consortium under Award No. W81XWH-10-2-0181, and through the Psychological Health and Traumatic Brain Injury Research Program under Award No. W81XWH-15-1-0632. Opinions, interpretations, conclusions and recommendations are those of the author and are not necessarily endorsed by the Department of Defense.

\ifCLASSOPTIONcaptionsoff
  \newpage
\fi

\bibliographystyle{IEEEtran}
\bibliography{mybib}

\end{document}